\documentclass[preprint, 3p, twocolumn]{elsarticle}
\biboptions{longnamesfirst}

\usepackage{natbib}
\usepackage{graphicx}% Include figure files
\usepackage{dcolumn}% Align table columns on decimal point
\usepackage{bm}% bold math
\usepackage{amsmath}
\usepackage{amsfonts}

\usepackage{amssymb}

 \usepackage{float}

\def\lb{\label}
\def\be{\begin{equation}}
\def\ee{\end{equation}}
\def\p{\hat{p}}
\def\q{\hat{q}}

\newcommand{\bea}{\begin{eqnarray}}
\newcommand{\eea}{\end{eqnarray}}

\newcommand{\ba}{\begin{array}} \newcommand{\ea}{\end{array}}

\begin{document}

\title{Conformal Triangles and Zig-Zag Diagrams}% Force line breaks with \\

\author[add1]{S.Derkachov}
\ead{derkach@pdmi.ras.ru}

\author[add1,add2]{A.P.Isaev}
\ead{isaevap@theor.jinr.ru}

\author[add1]{L.Shumilov}
\ead{la\_shum@mail.ru}

\address[add1]{St.Petersburg Department of the Steklov Mathematical Institute \it {of Russian Academy of Sciences,Fontanka 27, 191023, St.Petersburg, Russia.}}
\address[add2]{Bogoliubov  Laboratory of Theoretical Physics, {\it Joint Institute for Nuclear Research,
141980 Dubna, Russia.}}
%\address[3]{St.Petersburg Department of the Steklov Mathematical Institute \it {of Russian Academy of Sciences,Fontanka 27, 191023, St.Petersburg, Russia.}}


\date{\today}% It is always \today, today,

\begin{abstract}
   A convenient operator representation
for zig-zag four-point and
two-point planar Feynman diagrams relevant to
 the bi-scalar $D$-dimensional
 ''fishnet'' field theory is obtained. This representation
gives a possibility to evaluate exactly
diagrams of the zig-zag series in special cases.
In particular, we give a fairly simple  proof
of the Broadhurst-Kreimer conjecture about the values of
zig-zag multi-loop two-point diagrams which make a
significant contribution to the renormalization group $\beta$-function  in
the 4-dimensional $\phi^4$ theory.

\end{abstract}
\maketitle

\section{Introduction} 

The 4-dimensional $\phi^4$ field theory (and its multicomponent
generalizations) serves  the Brout-Englert-Higgs mechanism and thus
is an essential part of the Standard Model of particle physics. It was shown by explicit evaluation (in MS scheme) of the
Gell-Mann-Low $\beta$-function
in 4-dimensional $\phi^4$ theory
 that special Feynman diagrams (so-called zig-zag diagrams
 depicted in eqs. (\ref{zgzg3}) and (\ref{zgzg4}) below)
give 44\%, 46\% and 47\% of numerical
contributions, respectively, to the $3,4$ and $5$ loop orders of $\beta$
 \cite{BrKr}
  (see also \cite{Schnetz2, KP, BCK} and references therein for the
  explicit expression of the seven loop $\beta$-function in
  the $\phi^4_{D=4}$ theory). The $M$-loop zig-zag diagram
   gives the $(M+1)$-loop contribution to the $\beta$-function.
  The two-point function,
  which is represented by the $M$-loop zig-zag diagram (\ref{zgzg3}), (\ref{zgzg4}), has the general form
 \be
 \lb{BKr}
 G_2(x,y) = \frac{\pi^{2M}}{(x-y)^2} \; Z(M+1) \; ,
 \ee
 where $\pi^{2M}$ is the normalization
 factor (we discuss this factor at the end of the paper), $x,y \in \mathbb{R}^4$ and $Z(M+1)$ is a constant that contributes to the $\beta$-function in the $\phi^4_{D=4}$ theory
 in the $(M+1)$-loop order.
 The first nontrivial terms $Z(3)=6\zeta_3$ and $Z(4)=20\zeta_5$
 were analytically evaluated
 in \cite{ChTk1} and \cite{ChTk2}, respectively.
 The constant $Z(5)=\frac{441}{8}\zeta_7$
 of the zig-zag graph (\ref{zgzg3})
with 4 loops was calculated by D.Kazakov \cite{KD1} in 1983
(see also \cite{KD2}). The 5 loop zig-zag diagram
contribution $Z(6)=168\zeta_9$ to the $\beta$-function (in
6-loop order) was found
by D.Broadhurst \cite{Br1} in 1985 and confirmed by
N.Ussyukina \cite{Uss} in 1991.
Then D.Broadhurst and D.Kreimer in 1995 \cite{BrKr}
(see also \cite{BrKr2})
evaluated $Z(n)$ numerically up to $n=(M+1)=10$ loops,
and based on these data they
formulated a remarkable conjecture
that the constant $Z(M+1)$ is given by the following expression
\begin{align}\label{BK}
Z(M+1) &= 4 C_M
\sum\limits_{p=1}^{\infty}
\frac{(-1)^{(p-1)(M+1)}}{p^{2(M+1)-3}}\nonumber = \\=&\footnotesize
\left\{
\begin{array}{l}
4 \, C_M \, \zeta_{2M-1} \;\; {\rm for} \;\; M =2N+1 \, ,\\ [0.2cm]
4\, C_M \, (1 - 2^{2(1-M)}) \, \zeta_{2M-1} \;\;
{\rm for} \;\; M=2N \, ,
\end{array}
\right.
\end{align}
where $M$ is the number of loops in the zig-zag
diagrams (\ref{zgzg3}), (\ref{zgzg4})
and $C_M = \frac{1}{(M+1)} \binom{2M}{M}$ is the Catalan number.
The proof of the Broadhurst-Kreimer conjecture
was found in \cite{BS0,BS}. The proof of \cite{BS0,BS}
is based on the results of works \cite{Schnetz,Drum}.

We used a rather general approach to analytical
evaluation of the 2-point and 4-point zig-zag diagrams.
This approach leads to a fairly simple
 proof of the Broadhurst-Kreimer conjecture
 that is different from the proof of \cite{BS}.  Here
we make use of the operator formalism \cite{Isa}, \cite{Isa2} and
methods of \cite{GKK} based on the Euclidean multi-dimensional
conformal quantum field theories
(see \cite{DobMac},\cite{ToMiPe}, \cite{DolOsb1}, \cite{DolOsb2},
\cite{Osb3},  and references therein).

Our approach is partially inspired by the papers \cite{DKO,DO,DO1,DFO}
devoted to the  representation of separated variables for the
Basso--Dixon integrals \cite{BD}.
B.~Basso and L.~Dixon originally proposed in \cite{BD}
a nice explicit determinant formula for a
 family of Feynman diagrams in the $D=4$ fishnet conformal field 
 theories (CFT) \cite{KG,GGKK}. These diagrams can be constructed
by using the so-called graph-building
operator, such that the whole problem is essentially reduced
 to the problem of diagonalization
of this operator. A complete basis
of the corresponding eigenfunctions in the two-dimensional fishnet 
theory is constructed in \cite{DKO} by
using the methods of the theory of integrable spin chains. These results were generalized to the case of the
 four-dimensional fishnet theory in \cite{DO,DO1} and in \cite{DFO} to
the case of the $D$-dimensional fishnet theory \cite{KO}.
Typical Feynman diagrams in the fishnet CFT possess 
 special iterative structure and
are constructed by using various graph-building operators \cite{KG,GGKK,KO,GKK,BCF}.

 In this paper, we deduce the operator and integral presentations for
 four-point and two-point zig-zag  planar Feynman graphs relevant to
 bi-scalar $D$-dimensional fishnet theory. This gives
a possibility to compute exactly a special class of
 zig-zag  four-point and two-point multi-loop
Feynman diagrams in the $\phi^4_D$ theory. The eigenfunctions of
the corresponding graph-building operators are essentially simpler in
comparison with the Basso--Dixon diagrams and are in fact known
as special 3-point correlation functions (conformal triangles)
in CFT.

 \section{Operator formalism\label{opfor}}
 
 Consider $D$-dimensional Euclidean space $\mathbb{R}^D$
with the coordinates $x^\mu$ $(\mu=1,..., D)$ and
denote $\left(x\right)^{2\alpha}
 \equiv \left(x^2\right)^{\alpha}
= (\sum_\mu \, x^\mu \, x^\mu)^{\alpha} $.
Let $\{ \q^\mu_a , \; \p^\nu_b \}$ $(a,b=1,...,n)$
be Hermitian generators
of the algebra
 ${\cal H}^{(n)}=\sum_{a=1}^n {\cal H}_a$
consisting of n copies of the $D$-dimensional
Heisenberg algebras ${\cal H}_a$
\be
\lb{gr001}
[\q^\mu_a , \, \p^\nu_b ] = i \, \delta^{\mu \nu} \, \delta_{ab} \; .
\ee
We introduce states $|x_a \rangle$ and $|k_a \rangle$
which respectively diagonalize the generators
 $\q^\mu_a$ and $\p^\nu_a$ of the subalgebras
  ${\cal H}_a \subset {\cal H}^{(n)}$
\be
\lb{gr002}
\q^\mu_a |x_a \rangle = x^\mu_a |x_a \rangle \; , \;\;\;\;\;
\p^\mu_a |k_a \rangle = k^\mu_a |k_a \rangle \; ,
\ee
and form the basis in the space $V_a$, where the subalgebra
${\cal H}_a$ acts. The whole algebra ${\cal H}^{(n)}$ acts in
the space $V_1 \otimes \cdots \otimes V_n$ with the basis
elements $|x_1,...,x_n \rangle :=
|x_1 \rangle \otimes \cdots \otimes  |x_n \rangle$.
We also introduce the dual states $\langle x_a |$
and $\langle k_a |$ such that the orthogonality and
completeness conditions are valid
\be
\lb{ortcom}
\begin{array}{c}
\langle x_a | x'_a \rangle = \delta^D (x_a-x'_a) , \;\;\;
\langle k_a | k'_a \rangle = \delta^D (k_a-k'_a) , \;\;\; \\ [0.2cm]
\int \! d^D \!x_a \; |x_a \rangle \langle x_a | = I_a =
\int \! d^D \!k_a \; |k_a \rangle \langle k_a | \, ,
\end{array}
\ee
and $I_a$ is the unit operator in $V_a$.
Relations (\ref{gr001}), (\ref{gr002}),
(\ref{ortcom}) are consistent if we have
\begin{align*}
k^\mu_a \langle x_a |& k_a \rangle =
\langle x_a | \p^\mu_a |k_a \rangle = -i \frac{\partial}{\partial x^\mu_a}
\langle x_a | k_a \rangle \;\;\; \Rightarrow\\&\Rightarrow \;\;\;
\langle x_a | k_a \rangle = \frac{1}{(2\pi)^{D/2}}
e^{i k^\mu_a x^\mu_a} \; ,
\end{align*}
where there are no summations over the repeated index $a$, and
the normalization constant $(2\pi)^{-D/2}$ is 
fixed by (\ref{ortcom}). Below we use 
 the operators $(\q_a)^{-2\alpha} = 
 (\sum_\mu \q_a^\mu\ \q_a^\mu)^{-\alpha}$ and
$(\p_a)^{-2\beta}= (\sum_\mu \p_a^\mu\ \p_a^\mu)^{-\beta}$
with non-integer powers $\alpha$ and $\beta$.
These operators are understood as integral operators defined
 via their integral kernels
 $\langle x | \left(\q\right)^{-2\alpha} |y \rangle =
 \left(x\right)^{-2\alpha} \delta^D(x-y)$ and
\begin{align*}
    \langle x| \frac{1}{(\p)^{2\beta} } |y\rangle &=
\int d^D k \langle x| \frac{1}{(\p)^{2\beta} }|k \rangle
\langle k  |y\rangle =
  \nonumber\\ &\int  \, \frac{d^D k}{(2 \pi)^{D} } \,
 \frac{e^{i k ( x - y)}}{(k)^{2\beta} }=
\frac{a(\beta)}{\left(x-y\right)^{2\beta'} } \; ,
\end{align*}
\be
\lb{gr3}
\begin{array}{c}
\displaystyle
a(\beta):=
 \frac{1}{2^{2 \beta} \pi^{D/2}} \,
 \frac{\Gamma(\beta') }{ \Gamma(\beta) }
 \; , \;\;\;\;\; \beta' := D/2 - \beta \; .
 \end{array}
\ee
Now we consider the case of the algebra
${\cal H}^{(2)}= {\cal H}_1 + {\cal H}_2$
and introduce
\be
 \lb{qqd}
\hat{Q}_{12}^{(\beta)} : = \dfrac{1}{a(\beta)} \;
{\cal P}_{12} \; (\p_1)^{-2\beta} \; (\hat{q}_{12})^{-2\beta} \; ,
\ee
where $\hat{q}_{12}^{\mu} = \hat{q}_{1}^{\mu}
- \hat{q}_{2}^{\mu}$ and
${\cal P}_{12}$ is the permutation
\be
\nonumber
 {\cal P}_{12} \, \q_1 = \q_2 \, {\cal P}_{12} \; , \;\;\;\;
 {\cal P}_{12} \, \p_1 = \p_2 \, {\cal P}_{12} \; , \;\;\;\;
\ee
\be
\lb{perm}
 {\cal P}_{12}| x_1,x_2 \rangle  =
 | x_2,x_1 \rangle \; , \;\;\;\; ({\cal P}_{12})^2 = I \; .
\ee
  We depict the kernel
  of the operator (\ref{qqd}) as following

  \unitlength=4.5mm
\begin{picture}(25,4)(0,0)

\put(0.2,3){\footnotesize $x_1$}
\put(3.1,1.9){\tiny $\beta$}
\put(1.3,1.7){\tiny $\beta'$}
\put(0.2,1){\footnotesize $x_2$}
\put(3.2,0.8){\footnotesize $y_2$}
\put(3.2,3){\footnotesize $y_1$}

\put(3,1){\line(0,1){2}}
 \put(3,3){\line(-1,-1){2}}
 \multiput(3,1)(-0.18,0.18){12}{\circle*{0.1}}

 \put(4,1.8){$=$}
 %%%%%%%%%%%%%%%%%%%%%%%%%

 \put(5.1,3){\footnotesize $x_1$}
\put(8.1,1.9){\tiny $\beta$}
\put(6.7,0.5){\tiny $\beta'$}
\put(5.2,0.8){\footnotesize $x_2$}
\put(8.2,0.8){\footnotesize $y_1$}
\put(8.2,3){\footnotesize $y_2$}

\put(8,1){\line(0,1){2}}
 \put(8,1){\line(-1,0){2}}
 \multiput(8,3)(-0.24,0){10}{\circle*{0.1}}
%%%%%%%%%%%%%%%%%%%%

\put(9.5,1.8){\footnotesize $= \; \langle x_1,x_2 |
\hat{Q}_{12}^{(\beta)}  | y_1,y_2 \rangle \, =$}
\put(1, -1){$=
\, \frac{1}{a(\beta)} \, \langle x_1,x_2 |\;
{\cal P}_{12} \,(\p_1)^{-2\beta} \,
(\hat{q}_{12})^{-2\beta} \, | y_1,y_2 \rangle \, = $}

\put(3,-3){$= \;\;
\frac{1}{(x_2-y_1)^{2\beta'} \, (y_1-y_2)^{2\beta}} \; \delta^D(x_1-y_2)$ ,}

\end{picture}

\vspace{-0.9cm}
\be
\lb{figQD}
{}
\ee

\vspace{0.4cm}

\noindent
where

 \unitlength=6mm
\begin{picture}(25,1.5)

 %%%%%%%%%%%%%%%%%%%%%%%%%

\put(1,0.8){\footnotesize $x_1$}
\put(4.4,0.8){\footnotesize $x_2$}

\multiput(4,1)(-0.25,0){9}{\circle*{0.1}}

\put(5.6,0.8){$= \;\; \delta^D(x_1-x_2)$ ,}

 %%%%%%%%%%%%%%%%%%%%%%%%%

\put(1,-1){\footnotesize $x_1$}
\put(4.4,-1){\footnotesize $x_2$}
\put(2.9, -0.8){\footnotesize $\beta$}

\put(2,-1){\line(1,0){2}}
 %\multiput(16,1)(-0.25,0){9}{\circle*{0.1}}

\put(5.6,-1){$= \;\; (x_1-x_2)^{-2\beta}$ .}

\end{picture}
\vspace{0.8cm}

\noindent
The 4-dimensional analog of the kernel (\ref{figQD})
(for $\beta=1$) was
considered in \cite{GKK} and denoted there as $H_{1\!\!1}$.
Note that $Q_{12}^{(\beta)}$ is the graph building operator
for the planar zig-zag Feynman graphs: \\
 %for even loops

%\begin{widetext}
\noindent
for even loops

\unitlength=4mm
\begin{picture}(25,3.5)(1,1)

 \put(4.6,1.9){\tiny $\beta$}
\put(3.5,1.8){\tiny $\beta$}
\put(8.6,1.9){\tiny $\beta$}
\put(7.5,1.8){\tiny $\beta$}

 \put(3.8,3.2){\tiny $\beta'$}
 \put(3.8,0.5){\tiny $\beta'$}
  \put(5.8,3.2){\tiny $\beta'$}
 \put(5.8,0.5){\tiny $\beta'$}
 \put(7.8,0.5){\tiny $\beta'$}
 \put(8,3.2){\tiny $\beta'$}

\put(2.1,3){\footnotesize $x_1$}
\put(2.1,1){\footnotesize $x_2$}
\put(5,1){\line(0,1){2}}
\put(3,1){\line(1,0){2}}
\put(3,3){\line(1,0){2}}
 \put(5,1){\line(-1,1){2}}

 %\put(3,1){\line(0,1){2}}
 %\put(2.85,2.85){$\bullet$}
 %\put(16.85,0.85){$\bullet$}
%%%%%%%%%%%%%%%%%%%%%%%%%%%%

 \put(6.6,1.9){\tiny $\beta$}
 \put(5.5,1.8){\tiny $\beta$}

\put(17.2,0.8){\footnotesize $y_2$}
\put(17.2,3){\footnotesize $y_1$}
\put(7,1){\line(0,1){2}}
\put(5,1){\line(1,0){2}}
\put(5,3){\line(1,0){2}}
 \put(7,1){\line(-1,1){2}}
 \put(4.85,0.8){$\bullet$}
\put(4.85,2.8){$\bullet$}

 %%%%%%%%%%%%%%%%%%%%%%%%%%%%%%

 \put(6.85,0.8){$\bullet$}
\put(6.85,2.8){$\bullet$}
 \put(9,1){\line(0,1){2}}
\put(7,1){\line(1,0){2}}
\put(7,3){\line(1,0){2}}
 \put(9,1){\line(-1,1){2}}

\put(8.85,0.8){$\bullet$}
\put(8.85,2.8){$\bullet$}
%%%%%%%%%%%%%%%%%%%%%%%%%%%%%%
 \put(9.8,2){$. \; .\; . \; . \; .\; . $}
%%%%%%%%%%%%%%%%%%%%%%%%%%%%%%

 \put(13.8,0.5){\tiny $\beta'$}
 \put(14,3.2){\tiny $\beta'$}

\put(15,1){\line(0,1){2}}
\put(13,1){\line(1,0){2}}
\put(13,3){\line(1,0){2}}
 \put(15,1){\line(-1,1){2}}

%%%%%%%%%%%%%%%%%%%%%%%%%%%%

\put(15.8,0.5){\tiny $\beta'$}
\put(14.1,1.9){\tiny $\beta$}
\put(16.1,1.9){\tiny $\beta$}
\put(16,3.2){\tiny $\beta'$}

 %\put(17,1){\line(0,1){2}}
\put(15,1){\line(1,0){2}}
\put(15,3){\line(1,0){2}}
 \put(17,1){\line(-1,1){2}}
 \put(14.85,0.8){$\bullet$}
\put(14.85,2.8){$\bullet$}

\put(18.5,2){\footnotesize $=$}
\put(4,-2){\footnotesize
$= \;\;\;
\langle x_1,x_2 | (\hat{Q}_{12}^{(\beta)})^{2N}| y_1,y_2 \rangle
(y_1-y_2)^{2\beta} = $}

\end{picture}

\vspace{-1cm}

\be
\lb{zgzg1}
{}
\ee

\vspace{0.5cm}

\unitlength=1.8mm
\begin{picture}(4,12)(-4,-1)

%%%%%%%%%%%%%%%%%%%%%%%%%%%%

\put(-2,2.7){$=$}

\put(0.3,6.2){\footnotesize $x_1$}
\put(0.3,0.2){\footnotesize $x_2$}

%%%%%%%%%%%%Pervaya petlja%%%%%%%%%%%%%%%%

\qbezier(4.2,2.3)(5.1,-1.6)(6,2.5)

 \qbezier(6,2.5)(6.2,6.1)(2.5,5.7)

 \qbezier(4.2,2.3)(3.9,3.5)(5.2,4.6)

 %%%%%%%%%%%%%%%%%Vtoraya petlja%%%%%%%%%%%%%%%%%%%

 \qbezier(8,2.3)(8.9,-1.6)(9.8,2.5)
 \qbezier(9.8,2.5)(9.8,7)(6.3,5)

 \qbezier(8,2.3)(7.7,3.5)(9,4.6)

%%%%%%%%%%%%tret'ja petlja%%%%%%%%%%%%%%%%

 \qbezier(11.5,2.3)(12.4,-1.6)(13.3,2.5)
 \qbezier(13.3,2.5)(13.3,7)(9.8,5)

 \qbezier(11.5,2.3)(11.2,3.5)(12.5,4.6)

 %%%%%%%%%%%%%%%%%%Poslednie petli%%%%%%%%%%%%%%%%%%%%%%%%
 %%%%%%%%%%%%%%%%%%%%%%%%%%%%%%%%%%%%%%%%

  \put(14.5,3){$. \; .\; . \; . \; .\; . $}

 %%%%%%%%%%%%%%%%%4 petlja%%%%%%%%%%%%%%%%%%%

 \qbezier(22,2.3)(22.9,-1.6)(23.8,2.5)
 \qbezier(23.8,2.5)(23.8,7)(20.3,5)

 \qbezier(22,2.3)(21.7,3.5)(23,4.6)

%%%%%%%%%%%%5 petlja%%%%%%%%%%%%%%%%

 \qbezier(25.5,2.3)(26.4,-1.6)(27.3,2.5)
 \qbezier(27.3,2.5)(27.3,7)(23.8,5)

 \qbezier(25.5,2.3)(25.2,3.5)(26.5,4.6)

 \qbezier(27.5,5.1)(29.1,6.4)(30.2,4.8)

 \qbezier(30.2,4.8)(31,3)(30.1,0.4)

 \put(2.7,0.4){\line(1,0){27.5}}
 \put(3.4,5.7){\line(1,0){29}}

 %%%%%%%%%%%%%%%%%%%%%%%%%%%%%%%%%%
 \put(4.5,-0.2){$\bullet$}
 \put(8.4,-0.2){$\bullet$}
  \put(11.8,-0.2){$\bullet$}
  \put(22.4,-0.2){$\bullet$}
  \put(25.8,-0.2){$\bullet$}

 \put(7.5,5.2){$\bullet$}
  \put(11,5.2){$\bullet$}
  \put(21.6,5.2){$\bullet$}
  \put(25,5.2){$\bullet$}
   \put(28.4,5.2){$\bullet$}

 \put(33,6.2){\footnotesize $y_1$}
\put(30.8,0){\footnotesize $y_2$}

\put(36,3){\bf ;}
 %%%%%%%%%%%%%%%%%Primery%%%%%%%%%%%%%%%%%%
 %\qbezier(42,2.5)(42,3.5)(42.9,4.4)
 %\qbezier(43.8,3.5)(45.1,-1.6)(46,2.5)
 %\qbezier(12.7,4.1)(15.1,-0.6)(15.6,2.5)
 %\qbezier(44.5,2.5)(45.5,5)(46,2.5)
 %%%%%%%%%%%%%%%%%%%%%%%%%%

\end{picture}

\noindent
for odd loops

\unitlength=4mm
\begin{picture}(25,4)(2,1)

\put(2.2,3.2){\footnotesize $x_1$}
\put(2.1,0.6){\footnotesize $x_2$}

 %\put(4.5,3.4){\footnotesize $z_1$}
 %\put(4.7,0.5){\footnotesize $z_2$}
 %\multiput(5,1)(-0.25,0){9}{\circle*{0.13}}

\put(5,1){\line(0,1){2}}
\put(3,3){\line(1,0){2}}
 \put(5,1){\line(-1,1){2}}
 \put(3,1){\line(1,0){2}}

\put(3.8,0.5){\tiny $\beta'$}
 \put(4,3.2){\tiny $\beta'$}
 \put(5.8,0.5){\tiny $\beta'$}
 \put(6,3.2){\tiny $\beta'$}
 \put(7.8,0.5){\tiny $\beta'$}
 \put(8,3.2){\tiny $\beta'$}

 \put(4.6,1.9){\tiny $\beta$}
\put(3.5,1.8){\tiny $\beta$}
 \put(6.6,1.9){\tiny $\beta$}
\put(5.5,1.8){\tiny $\beta$}
 \put(8.6,1.9){\tiny $\beta$}
\put(7.5,1.8){\tiny $\beta$}
%%%%%%%%%%%%%%%%%%%%%%%%%%%%

\put(7,1){\line(0,1){2}}
\put(5,1){\line(1,0){2}}
\put(5,3){\line(1,0){2}}
 \put(7,1){\line(-1,1){2}}
 \put(4.85,0.8){$\bullet$}
\put(4.85,2.8){$\bullet$}

 %%%%%%%%%%%%%%%%%%%%%%%%%%%%%%

 \put(6.85,0.8){$\bullet$}
\put(6.85,2.8){$\bullet$}
 \put(9,1){\line(0,1){2}}
\put(7,1){\line(1,0){2}}
\put(7,3){\line(1,0){2}}
 \put(9,1){\line(-1,1){2}}

\put(8.85,0.8){$\bullet$}
\put(8.85,2.8){$\bullet$}
%%%%%%%%%%%%%%%%%%%%%%%%%%%%%%
 \put(10,2){$. \; .\; . \; . $}
%%%%%%%%%%%%%%%%%%%%%%%%%%%%%%

\put(14,1){\line(0,1){2}}
\put(12,1){\line(1,0){2}}
\put(12,3){\line(1,0){2}}
 \put(14,1){\line(-1,1){2}}

%%%%%%%%%%%%%%%%%%%%%%%%%%%%

\put(14.8,0.5){\tiny $\beta'$}
\put(15,3.2){\tiny $\beta'$}
\put(12.9,3.2){\tiny $\beta'$}
\put(12.7,0.5){\tiny $\beta'$}
\put(16.9,0.5){\tiny $\beta'$}
\put(16.1,1.9){\tiny $\beta$}
\put(14.6,1.7){\tiny $\beta$}
\put(13.6,1.9){\tiny $\beta$}
\put(12.6,1.7){\tiny $\beta$}

\put(16,1){\line(0,1){2}}
\put(14,1){\line(1,0){2}}
\put(14,3){\line(1,0){2}}
 \put(16,1){\line(-1,1){2}}
  %\put(11.85,0.8){$\bullet$}
  %\put(11.85,2.8){$\bullet$}
 \put(13.85,0.8){$\bullet$}
\put(13.85,2.8){$\bullet$}

\put(16,1){\line(1,0){2}}
 \put(15.85,0.8){$\bullet$}
\put(18.2,0.8){\footnotesize $y_1$}
\put(16.2,3){\footnotesize $y_2$}

\put(19.5, 2){\footnotesize $=$}
\put(4,-2){\footnotesize $= \;
\langle x_1,x_2 |
(\hat{Q}_{12}^{(\beta)})^{2N+1}| y_1,y_2 \rangle
(y_1-y_2)^{2\beta} =$}

\end{picture}
\vspace{-1cm}

\be
\lb{zgzg2}
{}
\ee

\vspace{0.5cm}

\unitlength=1.8mm
\begin{picture}(4,10)(-4,-1)

%%%%%%%%%%%%%%%%%%%%%%%%%%%%

\put(-2,2.7){$=$}

\put(0.3,6.2){\footnotesize $x_1$}
\put(0.3,0.2){\footnotesize $x_2$}

%%%%%%%%%%%%Pervaya petlja%%%%%%%%%%%%%%%%

\qbezier(4.2,2.3)(5.1,-1.6)(6,2.5)

 \qbezier(6,2.5)(6.2,6.1)(2.5,5.7)

 \qbezier(4.2,2.3)(3.9,3.5)(5.2,4.6)

 %%%%%%%%%%%%%%%%%Vtoraya petlja%%%%%%%%%%%%%%%%%%%

 \qbezier(8,2.3)(8.9,-1.6)(9.8,2.5)
 \qbezier(9.8,2.5)(9.8,7)(6.3,5)

 \qbezier(8,2.3)(7.7,3.5)(9,4.6)

%%%%%%%%%%%%tret'ja petlja%%%%%%%%%%%%%%%%

 \qbezier(11.5,2.3)(12.4,-1.6)(13.3,2.5)
 \qbezier(13.3,2.5)(13.3,7)(9.8,5)

 \qbezier(11.5,2.3)(11.2,3.5)(12.5,4.6)

 %%%%%%%%%%%%%%%%%%Poslednie petli%%%%%%%%%%%%%%%%%%%%%%%%
 %%%%%%%%%%%%%%%%%%%%%%%%%%%%%%%%%%%%%%%%

  \put(14.5,3){$. \; .\; . \; . \; .\; . $}

 %%%%%%%%%%%%%%%%%4 petlja%%%%%%%%%%%%%%%%%%%

 \qbezier(22,2.3)(22.9,-1.6)(23.8,2.5)
 \qbezier(23.8,2.5)(23.8,7)(20.3,5)

 \qbezier(22,2.3)(21.7,3.5)(23,4.6)

%%%%%%%%%%%%5 petlja%%%%%%%%%%%%%%%%

 \qbezier(25.5,2.3)(26.4,-1.6)(27.3,2.5)
 \qbezier(27.3,2.5)(27.3,7)(23.8,5)

 \qbezier(25.5,2.3)(25.2,3.5)(26.5,4.6)

 \qbezier(27.3,5)(28.5,5.8)(29.5,5.7)

 \put(2.7,0.4){\line(1,0){27.5}}
 \put(3.4,5.7){\line(1,0){26}}

 %%%%%%%%%%%%%%%%%%%%%%%%%%%%%%%%%%
 \put(4.5,-0.2){$\bullet$}
 \put(8.4,-0.2){$\bullet$}
  \put(11.8,-0.2){$\bullet$}
  \put(22.4,-0.2){$\bullet$}
  \put(25.8,-0.2){$\bullet$}

 \put(7.5,5.2){$\bullet$}
  \put(11,5.2){$\bullet$}
  \put(21.6,5.2){$\bullet$}
  \put(25,5.2){$\bullet$}
   %\put(28.4,5.2){$\bullet$}

 \put(30,6.2){\footnotesize $y_2$}
\put(30.8,0){\footnotesize $y_1$}

\put(36,3){\bf .}
 %%%%%%%%%%%%%%%%%Primery%%%%%%%%%%%%%%%%%%
 %\qbezier(42,2.5)(42,3.5)(42.9,4.4)
 %\qbezier(43.8,3.5)(45.1,-1.6)(46,2.5)
 %\qbezier(12.7,4.1)(15.1,-0.6)(15.6,2.5)
 %\qbezier(44.5,2.5)(45.5,5)(46,2.5)
 %%%%%%%%%%%%%%%%%%%%%%%%%%

\end{picture}
%\end{widetext}

\noindent
Here the bold face vertices denote the integration over
$\mathbb{R}^D$. We stress that the Feynman integrals, which correspond
to the matrix elements (\ref{zgzg1}), (\ref{zgzg2}) represent
the contribution to the 4-point correlation functions in the bi-scalar
$D$-dimensional ''fishnet'' theory (see \cite{GKK} and references therein).
For clarity, in the right hand sides
of (\ref{zgzg1}) and (\ref{zgzg2}), we present the zig-zag diagrams
in the form of spiral graphs having the cylindrical topology \cite{GKK}.
We also stress that integral kernels
(\ref{zgzg1}) and (\ref{zgzg2}),
 in the case of $D=4$ and $\beta=1$, contribute to
 Green's functions of the standard $\phi^4$ field theory.

The next important
statement is that $Q_{12}^{(\beta)}$, given in (\ref{qqd}),
is the graph building operator
for the integrals of the planar zig-zag two-point Feynman graphs: \\

\noindent
for even loops

\unitlength=4mm
\begin{picture}(25,3.5)(2,1)

\put(2.6,1.9){\tiny $\beta$}
 \put(4.6,1.9){\tiny $\beta$}
\put(3.5,1.8){\tiny $\beta$}
\put(8.6,1.9){\tiny $\beta$}
\put(7.5,1.8){\tiny $\beta$}

 \put(3.8,3.2){\tiny $\beta'$}
 \put(3.8,0.5){\tiny $\beta'$}
  \put(5.8,3.2){\tiny $\beta'$}
 \put(5.8,0.5){\tiny $\beta'$}
 \put(7.8,0.5){\tiny $\beta'$}
 \put(8,3.2){\tiny $\beta'$}

\put(2.1,3){\footnotesize $x_1$}
\put(2.1,1){\footnotesize $x_2$}
\put(5,1){\line(0,1){2}}
\put(3,1){\line(1,0){2}}
\put(3,3){\line(1,0){2}}
 \put(5,1){\line(-1,1){2}}

 %\put(3,1){\line(0,1){2}}
 %\put(2.85,2.85){$\bullet$}
 %\put(16.85,0.85){$\bullet$}
%%%%%%%%%%%%%%%%%%%%%%%%%%%%

 \put(6.6,1.9){\tiny $\beta$}
 \put(5.5,1.8){\tiny $\beta$}

\put(17.4,0.8){\footnotesize $y_2$}
\put(17.3,3){\footnotesize $y_1$}
\put(7,1){\line(0,1){2}}
\put(5,1){\line(1,0){2}}
\put(5,3){\line(1,0){2}}
 \put(7,1){\line(-1,1){2}}
 \put(4.85,0.8){$\bullet$}
\put(4.85,2.8){$\bullet$}

 %%%%%%%%%%%%%%%%%%%%%%%%%%%%%%

 \put(6.85,0.8){$\bullet$}
\put(6.85,2.8){$\bullet$}
 \put(9,1){\line(0,1){2}}
\put(7,1){\line(1,0){2}}
\put(7,3){\line(1,0){2}}
 \put(9,1){\line(-1,1){2}}

\put(8.85,0.8){$\bullet$}
\put(8.85,2.8){$\bullet$}
%%%%%%%%%%%%%%%%%%%%%%%%%%%%%%
 \put(9.8,2){$. \; .\; . \; . \; .\; . $}
%%%%%%%%%%%%%%%%%%%%%%%%%%%%%%

 \put(13.8,0.5){\tiny $\beta'$}
 \put(14,3.2){\tiny $\beta'$}

\put(15,1){\line(0,1){2}}
\put(13,1){\line(1,0){2}}
\put(13,3){\line(1,0){2}}
 \put(15,1){\line(-1,1){2}}

%%%%%%%%%%%%%%%%%%%%%%%%%%%%

\put(15.8,0.5){\tiny $\beta'$}
\put(16,3.2){\tiny $\beta'$}
\put(14.1,1.9){\tiny $\beta$}
\put(16.1,1.9){\tiny $\beta$}
\put(15.1,1.9){\tiny $\beta$}
\put(17.1,1.9){\tiny $\beta$}

\put(15,1){\line(1,0){2}}
\put(15,3){\line(1,0){2}}
 \put(17,1){\line(-1,1){2}}
 \put(14.85,0.8){$\bullet$}
\put(14.85,2.8){$\bullet$}

%%%%%%%%%%%%%%%%%%%%%%%%%%%%%%%%%
 \put(17,1){\line(0,1){2}}
 \put(3,1){\line(0,1){2}}
 \put(16.85,0.8){$\bullet$}
\put(2.85,2.8){$\bullet$}
 %%%%%%%%%%%%%%%%%%%%%%%%%%%

\put(18.5, 2){\footnotesize $=$}
\put(2.5,-2){\footnotesize
$\;\; = \;\; \int d^D x_1 d^D y_2
\dfrac{\langle x_1,x_2 | (\hat{Q}_{12}^{(\beta)})^{2N}| y_1,y_2 \rangle} {(x_1-x_2)^{2\beta}}\;$ ;}

\end{picture}

%\vspace{0cm}

\be
\lb{zgzg3}
{}
\ee

\vspace{0.5cm}

\noindent
for odd loops

\unitlength=4mm
\begin{picture}(25,4)(2,1)

\put(2.2,3.2){\footnotesize $x_1$}
\put(2.1,0.6){\footnotesize $x_2$}

\put(5,1){\line(0,1){2}}
\put(3,3){\line(1,0){2}}
 \put(5,1){\line(-1,1){2}}
 \put(3,1){\line(1,0){2}}

\put(3.8,0.5){\tiny $\beta'$}
 \put(4,3.2){\tiny $\beta'$}
 \put(5.8,0.5){\tiny $\beta'$}
 \put(6,3.2){\tiny $\beta'$}
 \put(7.8,0.5){\tiny $\beta'$}
 \put(8,3.2){\tiny $\beta'$}

 \put(4.6,1.9){\tiny $\beta$}
\put(3.5,1.8){\tiny $\beta$}
 \put(6.6,1.9){\tiny $\beta$}
\put(5.5,1.8){\tiny $\beta$}
 \put(8.6,1.9){\tiny $\beta$}
\put(7.5,1.8){\tiny $\beta$}
%%%%%%%%%%%%%%%%%%%%%%%%%%%%

\put(7,1){\line(0,1){2}}
\put(5,1){\line(1,0){2}}
\put(5,3){\line(1,0){2}}
 \put(7,1){\line(-1,1){2}}
 \put(4.85,0.8){$\bullet$}
\put(4.85,2.8){$\bullet$}

 %%%%%%%%%%%%%%%%%%%%%%%%%%%%%%

 \put(6.85,0.8){$\bullet$}
\put(6.85,2.8){$\bullet$}
 \put(9,1){\line(0,1){2}}
\put(7,1){\line(1,0){2}}
\put(7,3){\line(1,0){2}}
 \put(9,1){\line(-1,1){2}}

\put(8.85,0.8){$\bullet$}
\put(8.85,2.8){$\bullet$}
%%%%%%%%%%%%%%%%%%%%%%%%%%%%%%
 \put(10,2){$. \; .\; . \; . $}
%%%%%%%%%%%%%%%%%%%%%%%%%%%%%%

\put(14,1){\line(0,1){2}}
\put(12,1){\line(1,0){2}}
\put(12,3){\line(1,0){2}}
 \put(14,1){\line(-1,1){2}}

%%%%%%%%%%%%%%%%%%%%%%%%%%%%

\put(14.8,0.5){\tiny $\beta'$}
\put(15,3.2){\tiny $\beta'$}
\put(12.9,3.2){\tiny $\beta'$}
\put(12.7,0.5){\tiny $\beta'$}
\put(16.9,0.5){\tiny $\beta'$}
\put(16.1,1.9){\tiny $\beta$}
\put(17.2,2){\tiny $\beta$}
\put(14.6,1.7){\tiny $\beta$}
\put(13.6,1.9){\tiny $\beta$}
\put(12.6,1.7){\tiny $\beta$}

\put(16,1){\line(0,1){2}}
\put(14,1){\line(1,0){2}}
\put(14,3){\line(1,0){2}}
 \put(16,1){\line(-1,1){2}}
  %\put(11.85,0.8){$\bullet$}
  %\put(11.85,2.8){$\bullet$}
 \put(13.85,0.8){$\bullet$}
\put(13.85,2.8){$\bullet$}

\put(16,1){\line(1,0){2}}
 \put(15.85,0.8){$\bullet$}
\put(18.2,0.8){\footnotesize $y_1$}
\put(16.4,3){\footnotesize $y_2$}

%%%%%%%%%%%%%%%%%%%%%%%%%%%%%%%%%
 \put(3,1){\line(0,1){2}}
  \put(18,1){\line(-1,1){2}}
 \put(15.85,2.8){$\bullet$}
\put(2.85,2.8){$\bullet$}
 %%%%%%%%%%%%%%%%%%%%%%%%%%%

\put(19,2){\footnotesize =}
\put(2,-2){\footnotesize
$\; = \;\; \int d^D x_1 d^D y_2
\dfrac{\langle x_1,x_2 | (\hat{Q}_{12}^{(\beta)})^{2N+1}| y_1,y_2 \rangle} {(x_1-x_2)^{2\beta}}\,$ .}

\end{picture}

\be
\lb{zgzg4}
{}
\ee

\vspace{0.6cm}

\noindent
In the next sections, we use the operator representations
(\ref{zgzg1}) -- (\ref{zgzg4}) to evaluate
exactly the corresponding class of 2-point and 4-point
Feynman diagrams.

Finally, we note that the elements
 $H_\beta : = {\cal P}_{12} \, \hat{Q}_{12}^{(\beta)}\equiv
 (\p_1)^{-2\beta}  (\hat{q}_{12})^{-2\beta}$ form
 a commutative set of operators $[H_\alpha , \;  H_\beta]=0$
 $(\forall \alpha,\beta)$. This fact can be easily demonstrated
 by means of the operator version \cite{Isa} of the
 star-triangle relation.
 Special forms of the elements $H_\beta$
 were used in \cite{Isa} as graph-building operators for ladder
 diagrams.

\section{Eigenfunctions for the
graph building operator $\hat{Q}_{12}$.
Scalar product and completeness}
To find eigenvectors for the
graph building operator (\ref{qqd}) we consider
the standard 3-point correlation function
of three fields ${\cal O}_{\Delta_1}$, ${\cal O}_{\Delta_2}$
and ${\cal O}_{\Delta}^{\mu_1...\mu_n}$
 in a conformal field theory.
 Here  ${\cal O}_{\Delta_1}$ and ${\cal O}_{\Delta_2}$
 are two scalar fields with conformal dimensions
 $\Delta_1$ and $\Delta_2$, while ${\cal O}_{\Delta}^{\mu_1...\mu_n}$
 is a tensor field with conformal dimension
 $\Delta$. The single conformally-invariant
tensor structure of this correlation function
(up to a normalization) is well known
\begin{align}
 \lb{cor01}
 u^{\mu_1} \cdots u^{\mu_n}& \;
 \langle  {\cal O}_{\Delta_1}(y_1) \;
 {\cal O}_{\Delta_2}(y_2) \; {\cal O}_{\Delta}^{\mu_1...\mu_n}(y) \rangle\nonumber =\\ =
 &\frac{\bigl( \frac{(u,y-y_1)}{(y-y_1)^2} -
\frac{(u,y-y_2)}{(y-y_2)^2} \bigr)^{n}}{(y_1-y_2)^{2A}
(y-y_1)^{2A_1}(y-y_2)^{2A_2}} \; ,
\end{align}
 where
 $$
 A = \frac{1}{2}(\Delta_1+\Delta_2 -\Delta +n) \, , \;\;\;
 A_1 = \frac{1}{2}(\Delta_1+\Delta -\Delta_2 -n) \, , \;\;\;
 $$
 $$
 A_2 = \frac{1}{2}(\Delta_2+\Delta -\Delta_1 - n) \, .
 $$
 Here and below we make formulas concise by
 using an auxiliary complex vector $u \in \mathbb{C}^D$
 such that $(u,u) = u^\mu u^\mu = 0$.
 We need the special form of the 3-point function
 (\ref{cor01})
 when parameters $A,A_1,A_2$ are related
 to two numbers $\alpha \in \mathbb{C}$,
 $\beta \in \mathbb{R}$:
 \be
 \lb{cor00}
 \begin{array}{c}
 A = \alpha \; , \;\;\; A_1 = \alpha' := \frac{D}{2} - \alpha
 \; , \;\;\; \\[0.2cm] A_2 = (\alpha+ \beta)' = \frac{D}{2} -
 (\alpha + \beta) \; \Rightarrow \; \\ [0.2cm]
 \Delta_1 = \frac{D}{2} \; , \;\;\; \Delta_2 = \frac{D}{2}-\beta
 \; , \;\;\; \Delta = D-2\alpha-\beta+n \; ,
 \end{array}
 \ee
 %and transform the tensor field ${\cal O}_{\Delta}^{\mu_1...\mu_n}(y)$
 %into  momentum representation
 so we have
\begin{align}
 \lb{cor02}
 \langle y_1,y_2 |& \Psi_{\alpha,\beta}^{(n,u)}(y) \rangle \; \nonumber : = \;\\
 &=\frac{\bigl( \frac{(u,y-y_1)}{(y-y_1)^2} -
\frac{(u,y-y_2)}{(y-y_2)^2} \bigr)^{n}}{(y_1-y_2)^{2\alpha}
(y-y_1)^{2\alpha'}(y-y_2)^{2(\alpha+\beta)'}}  \; ,
\end{align}
 and, following the paper \cite{VPH}, we call this
 function as conformal triangle.
 It is a remarkable fact that
 $| \Psi_{\alpha,\beta}^{(n,u)}(y) \rangle =
 u^{\mu_1} \cdots u^{\mu_n}
 | \Psi_{\alpha,\beta}^{\mu_1...\mu_n}(y) \rangle$, defined
 in (\ref{cor02}), is the eigenvector for the
 graph building operator (\ref{qqd})
 \be
 \lb{cor03}
 \hat{Q}^{(\beta)}_{12} \;
 | \Psi_{\alpha,\beta}^{(n,u)}(y) \rangle = \tau(\alpha,\beta,n)
 \; | \Psi_{\alpha,\beta}^{(n,u)}(y) \rangle \; ,
 \ee
 with the eigenvalue
\be
 \lb{cor04}
 \tau(\alpha,\beta,n) = (-1)^n \,
\frac{\pi^{D/2} \Gamma(\beta) \Gamma(\alpha)
\Gamma((\alpha+\beta)'+n)}{ \Gamma(\beta') \Gamma(\alpha'+n) \Gamma(\alpha+\beta)}  \; .
 \ee
 An analogous statement,
 for $D=4$ and $\beta=1$, was announced 
 in \cite{GKK}.

  Note that with respect to the standard Hermitian
 scalar product
\begin{align}
 \lb{scpro}
 \langle \Psi | \Phi \rangle &=
 \int d^D x_1 \, d^D x_2 \,
 \langle \Psi | x_1,x_2 \rangle
 \langle  x_1,x_2| \Phi \rangle =\nonumber \\ &=
  \int d^4 x_1 d^4 x_2 \, \Psi^*(x_1,x_2) \,\Phi(x_1,x_2) \; ,
\end{align}
 the operator (\ref{qqd}) for $\beta \in \mathbb{R}$
 is Hermitian up to the equivalence transformation:
  \be
 \nonumber
 \lb{hermc}
 (\hat{Q}_{12}^{(\beta)})^\dagger =
 \frac{1}{a(\beta)} \; (\hat{q}_{12})^{-2\beta}  \; (\p_1)^{-2\beta}
 \; {\cal P}_{12} = U \, \hat{Q}_{12}^{(\beta)} \, U^{-1}\;\, ,
 \ee
 \be
 \;\;  U:= {\cal P}_{12} \, (\hat{q}_{12})^{-2\beta}=
 (\hat{q}_{12})^{-2\beta} \, {\cal P}_{12} \, .
 \ee
 %where\be\lb{uuu}
 %U:= {\cal P}_{12} \, \hat{q}_{12}^{-2\beta}=
 %\hat{q}_{12}^{-2\beta} \, {\cal P}_{12}
 %=\tilde{a}(\beta)\; \p_2^{2\beta} \, \hat{Q}_{12}^{(\beta)}\ee
  Therefore, we modify the scalar product (\ref{scpro})
 \begin{align}
 \nonumber
 \lb{scprod}
\langle \overline{\Psi}|\Phi\rangle :=
\langle \Psi| \, U \, |\Phi\rangle &=
\int d^4x_1 d^4x_2\frac{\Psi^{*}(x_2\,,x_1) \,\Phi(x_1\,,x_2)}{
(x_1-x_2)^{2\beta}} \; , \\\;\;\;\;\; &(\beta \equiv D-\Delta_1 - \Delta_2) \; ,
\end{align}
 and with respect to this scalar product
 the operator (\ref{qqd}) is Hermitian.
In (\ref{scprod}) a special conjugation for the vectors
was introduced
\be
\lb{nconj}
\langle \overline{\Psi}|:= \langle \Psi| \, U =
\langle \Psi | \, (\q_{12})^{-2\beta} \, {\cal P}_{12} \; ,
 \ee
 and the operator $U$ in (\ref{hermc}) plays the role of
 the metric in the space $V_1 \otimes V_2$.
  It is evident that the graph building operator (\ref{qqd}) commutes
 with
 \be
 \lb{qpd}
 \hat{\sf D} = \frac{i}{2} \sum_{a=1}^2 (\q_a \p_a + \p_a \q_a)
 + \frac{1}{2}(y^\mu \, \partial_{y^\mu}
 + \partial_{y^\mu} \, y^\mu) - \beta ,
 \ee
 which acts on the eigenvector
 $| \Psi_{\alpha,\beta}^{(n,u)}(y)\rangle$ as follows:
 \be
\lb{qpd2}
 \hat{\sf D} \;
 | \Psi_{\alpha,\beta}^{(n,u)}(y) \rangle =
 \Bigl(2\alpha+\beta-\frac{1}{2} D -n \Bigr) \;
 | \Psi_{\alpha,\beta}^{(n,u)}(y) \rangle \; ,
 \ee
 and, for $\beta \in \mathbb{R}$, satisfies
 $\hat{\sf D}^\dagger = - U \, \hat{\sf D}\, U^{-1}$. Thus,
 the operator $\hat{\sf D}$ is anti-Hermitian
 with respect to the scalar product (\ref{scprod}), and
 the corresponding condition on its eigenvalue gives
 \be
 \lb{cor05}
 2(\alpha^* + \alpha) = 2n +D-2\beta \;\;\;\; \Rightarrow \;\;\;\;
 \nonumber
 \ee
 \vspace{-0.8cm}
 \be
 \alpha = \frac{1}{2}\left(n + D/2 - \beta\right) - i \nu \; , \;\;\;\;
 \nu \in \mathbb{R} \; .
 \ee
 It is a remarkable fact that under this condition
  the eigenvalue (\ref{cor04}) is real
\begin{align}
 \lb{cor11}
 \tau(\alpha,\beta,n) =  (-1)^n \,
&\frac{\pi^{D/2} \Gamma(\beta) \; \Gamma(\frac{D}{4} + \frac{n}{2} -
\frac{\beta}{2} + i \nu) \,}{ \Gamma(\beta') \;
\Gamma(\frac{D}{4} + \frac{n}{2} +\frac{\beta}{2} + i \nu) \,
} \;
\nonumber\times\\&\times\frac{\Gamma(\frac{D}{4} + \frac{n}{2} -
\frac{\beta}{2} - i \nu)}{\Gamma(\frac{D}{4} + \frac{n}{2} +\frac{\beta}{2} - i \nu)},
\end{align}
 and the parameter $\Delta$ in (\ref{cor00})
 acquires the form $\Delta = \frac{D}{2} + 2 i \nu$.
 In view of this, we denote the eigenvector (\ref{cor02}) as
\be
 \lb{cor06}
 | \Psi_{\nu,\beta,y}^{(n,u)}\rangle :=
 | \Psi_{\alpha,\beta}^{(n,u)}(y) \rangle =
 u^{\mu_1}\cdots u^{\mu_n}| \Psi_{\alpha,\beta}^{\mu_1...\mu_n}(y) \rangle
  , \;\;\,
  \nonumber
  \ee
  \be\Psi_{\nu,\beta,y}^{(n,u)}(x_1\,,x_2) :=
 \langle x_1\,,x_2 | \Psi_{\nu,\beta,y}^{(n,u)} \rangle\, .
 \ee
 Since the eigenvalue (\ref{cor11}) is real (it is invariant under
 the transformation $\nu \to - \nu$), two eigenvectors
 $| \Psi_{\nu,\beta,x}^{(n,u)}\rangle$ and
 $| \Psi_{\lambda,\beta,y}^{(m,v)}\rangle$,
 having different eigenvalues (\ref{cor11})
 (e.g. $n \neq m$ and $\lambda \neq \pm \nu$), should be
 orthogonal to each other with respect to
 the scalar product (\ref{scprod}). Indeed, we have the following
 orthogonality condition
 for two conformal triangles (see, e.g.,
 \cite{DobMac}, \cite{ToMiPe}, \cite{DPPT}, \cite{GKK}):

\begin{align}\label{cor07}
\langle &\overline{\Psi_{\lambda,\beta,y}^{(m,v)}}|
\Psi_{\nu,\beta,x}^{(n,u)}\rangle =\nonumber\\ =
 &\int \, d^D x_1 \, d^D x_2 \; \langle \Psi_{\lambda,\beta,y}^{(m,v)}|U|
x_1 x_2 \rangle \langle x_1 x_2 |\Psi_{\nu,\beta,x}^{(n,u)}\rangle 
 = \nonumber\\ =&\int \, d^D x_1 \, d^D x_2
\frac{(\Psi_{\lambda,\beta,y}^{(m,v)}(x_2\,,x_1))^* \;
\Psi_{\nu,\beta,x}^{(n,u)}(x_1\,,x_2)}
{(x_1-x_2)^{2(D-\Delta_1-\Delta_2)}} 
=
\nonumber\\ =\,& C_1(n\,,\nu)\,\delta_{n m}\,\delta(\nu -\lambda)\,\delta^D(x-y)\,(u,v)^n + \nonumber\\ +\,
&C_2(n\,,\nu)\,\delta_{n m}\,\delta(\nu +\lambda)\,
\frac{\left((u,v)-2\frac{(u,x-y)(v,x-y)}{(x-y)^2}\right)^n}
{(x-y)^{2\left(D/2+2i\nu\right)}} ,
\end{align}

where $(u,v) = u^\mu v^\mu$,
$\;\beta = D-\Delta_1-\Delta_2 = \Delta_1-\Delta_2$ and
\begin{align}\lb{C1}
C_1(n\,,&\nu) =
\frac{(-1)^n\,2^{1-n}\,\pi^{3D/2 + 1}\,n!\,
}{
\Gamma\left(\frac{D}{2} + n\right)
\left(\left(\frac{D}{2}+n-1\right)^2 + 4\nu^2\right)\,}\times \nonumber\\
&\times\frac{\Gamma\left(2i\nu\right)\Gamma\left(-2i\nu\right)}{\Gamma\left(\frac{D}{2} + 2i - 1\nu\right)\Gamma\left(\frac{D}{2} - 2i\nu - 1\right)}
\end{align}
We note that the coefficient $C_1$ is independent of
$\beta$ and plays the important role as the inverse of
the Plancherel measure used
in the completeness condition (see below). In contrast to this,
the coefficient $C_2$ in (\ref{cor07})
depends on $\beta$, but the explicit form for $C_2$
 will not be important for us.
 %\begin{multline}\label{C2}
 %C_2(n\,,\nu) \stackrel{??}{=} 2 \pi\,
 %\pi^{D}\,\frac{n!}{2^n}\,
 %\frac{\Gamma\left(\frac{D}{4}-\frac{\Delta_1-\Delta_2}{2}+
 %\frac{n}{2}-i\nu\right)}
 %{\Gamma\left(\frac{D}{4}-\frac{\Delta_1-\Delta_2}{2}+
 %\frac{n}{2}+i\nu\right)}\,
 %\frac{\Gamma\left(\frac{D}{4}+\frac{\Delta_1-\Delta_2}{2}+
 %\frac{n}{2}-i\nu\right)}
 %{\Gamma\left(\frac{D}{4}+\frac{\Delta_1-\Delta_2}{2}+
 %\frac{n}{2}+i\nu\right)}\; \cdot\\
 %\frac{\Gamma\left(2i\nu\right)\Gamma\left(\frac{D}{2}+2i\nu-1+n\right)}
 %{\Gamma\left(\frac{D}{2}+n-2i\nu\right)
 %\Gamma\left(\frac{D}{2}+2i\nu-1\right)\Gamma\left(\frac{D}{2}+n\right)}
 %\end{multline}
 Respectively, completeness (or resolution of unity $I$)
 for the basis of eigenfunctions (\ref{cor06}) is~written~as~(see,~e.g.,\cite{DobMac}, \cite{ToMiPe}, \cite{DPPT}, \cite{GKK})
\begin{align}
    \label{cor08}
    I = &\sum_{n = 0}^{\infty}\int\limits_{0}^{\infty}\frac{d\nu}{C_1(n\,, \nu)}\int d^Dx |\Psi^{\mu_1\cdots\mu_n}_{\nu,\beta,x}\rangle\langle \overline{\Psi^{\mu_1\cdots\mu_n}_{\nu,\beta,x} }| \nonumber\\
     =&\sum_{n = 0}^{\infty}\int\limits_{0}^{\infty}\frac{d\nu}{C_1(n\,, \nu)}\int d^Dx |\Psi^{\mu_1\cdots\mu_n}_{\nu,\beta,x}\rangle\langle \Psi^{\mu_1\cdots\mu_n}_{\nu,\beta,x} |U.
\end{align}
%\begin{equation}
% \lb{cor08}
% \begin{array}{c}
% \displaystyle
% I = \sum_{n=0}^{\infty}
%\int_{0}^{\infty} \frac{d \nu}{C_1(n\,,\nu)} \int d^D x\,
%| \Psi^{\mu_1\cdots\mu_n}_{\nu,\beta,x}\rangle
%\langle \overline{\Psi^{\mu_1\cdots\mu_n}_{\nu,\beta,x} } |  = \\ [0.3cm]
%\displaystyle
%= \sum_{n=0}^{\infty}
%\int_{0}^{\infty} \frac{d \nu}{C_1(n\,,\nu)} \int d^D x\,
%| \Psi^{\mu_1\cdots\mu_n}_{\nu,\beta,x}\rangle
%\langle \Psi^{\mu_1\cdots\mu_n}_{\nu,\beta,x} |\, U \; .
 %\;\;\; \Rightarrow
% \end{array}
%\end{equation}
 Applying to this relation the vector $|y_1,y_2\rangle$
 from the right and the vector $\langle x_1,x_2|$ from the left
 and using the formulas
 $\langle \Psi^{\mu_1\cdots\mu_n}_{\nu,\beta,x} |x_1,x_2\rangle=
 (\Psi^{\mu_1\cdots\mu_n}_{\nu,\beta,x}(x_1,x_2))^*=
 \Psi^{\mu_1\cdots\mu_n}_{-\nu,\beta,x}(x_1,x_2)$,
 we write the resolution of unity (\ref{cor08})
 in terms of the integral
 kernels (see e.g. eq. (A.7) in \cite{GKK}).
\section{Four-point and two-point correlation
 functions for zig-zag diagrams}    
 Substitution of the resolution of unity (\ref{cor08}) into
 expressions (\ref{zgzg1}), (\ref{zgzg2}) for zig-zag 4-point
 Feynman graphs gives

\begin{align}\label{cor09}
 &G^{(M)}_4(x_1,x_2;y_1,y_2) = \nonumber\\
 =&\,\langle x_1,x_2 | \bigl(\hat{Q}_{12}^{(\beta)}\bigr)^{M}| y_1,y_2 \rangle
(y_1-y_2)^{2\beta} = \nonumber\\
=&\sum\limits_{n=0}^{\infty}
\int\limits_{0}^{\infty} \frac{d \nu}{C_1(n\,,\nu)} \int d^D x\,
\langle x_1,x_2 | \bigl(\hat{Q}_{12}^{(\beta)}\bigr)^{M}\times \nonumber\\ \times&|
\Psi^{\mu_1\cdots\mu_n}_{\nu,\beta,x}\rangle
\langle \Psi^{\mu_1\cdots\mu_n}_{\nu,\beta,x} |\, U  | y_1,y_2 \rangle
(y_1-y_2)^{2\beta}  = \nonumber\\
=& \sum\limits_{n=0}^{\infty}
\int\limits_{0}^{\infty} d \nu \,
\frac{\bigl(\tau(\alpha,\beta,n)\bigr)^M}{C_1(n\,,\nu)} \times\nonumber\\ \times&\int d^D x\,
\langle x_1,x_2 |\Psi^{\mu_1\cdots\mu_n}_{\nu,\beta,x}\rangle
\langle \Psi^{\mu_1\cdots\mu_n}_{\nu,\beta,x} |y_2,y_1 \rangle \, ,
\end{align}

where the integral over $x$ in the right-hand side
of (\ref{cor09}) is evaluated
in terms of conformal blocks \cite{DolOsb1}, \cite{DolOsb2}, \cite{DPPT}
(in the four-dimensional case, this integral was considered in detail
in \cite{GKK}).

Making use of the standard relations between the \linebreak 4-point zig-zag functions $G^{(M)}_4(x_1,x_2;y_1,y_2)$ constructed
 in (\ref{cor09})
and 2-point zig-zag functions
$G^{(M)}_2(x_2,y_1)$ (the graphs for these functions are presented
in (\ref{zgzg1}) -- (\ref{zgzg4})),
 we write explicit expressions
for the 2-point $M$-loop zig-zag diagrams as follows:

\begin{align}\label{cor12}
&G^{(M)}_2(x_2,y_1)
= \sum\limits_{n=0}^{\infty}
\int\limits_{0}^{\infty} d \nu
\frac{(\tau(\alpha,\beta,n))^M}{C_1(n\,,\nu)}
 \, \times\nonumber\\\times&\int d^D x_1 d^D y_2 \, d^D x\,
\frac{\langle x_1,x_2 |\Psi^{\mu_1\cdots\mu_n}_{\nu,\beta,x}\rangle
\langle \Psi^{\mu_1\cdots\mu_n}_{\nu,\beta,x} |y_2,y_1 \rangle}{
(x_1-x_2)^{2\beta}(y_1-y_2)^{2\beta}} = \nonumber\\
= &\frac{1}{(x_2-y_1)^{2\beta}} \frac{\Gamma(D/2-1)}{\Gamma(D-2)}
\sum\limits_{n=0}^{\infty}
\frac{(-1)^n \Gamma(n+D-2)}{2^{n} \Gamma(n+D/2-1)}\times \nonumber \\ \times&
\int\limits_{0}^{\infty} d \nu \;
\frac{\tau^{M+3}(\alpha,\beta,n)}{C_1(n\,,\nu)} \, ,
\end{align}

  where we apply the two-point master integral
\begin{multline}\label{cor13}
 \int d^D x_1 d^D y_2 \, d^D x\,
\frac{\langle x_1,x_2 |\Psi^{\mu_1\cdots\mu_n}_{\nu,\beta,x}\rangle
\langle \Psi^{\mu_1\cdots\mu_n}_{\nu,\beta,x} |y_2,y_1 \rangle}{
(x_1-x_2)^{2\beta}(y_1-y_2)^{2\beta}} = \\
 = \frac{(-1)^n \Gamma(n+D-2)\Gamma(D/2-1)}{2^{n} \Gamma(n+D/2-1)\Gamma(D-2)}
\, \frac{\tau^3(\alpha,\beta,n)}{(x_2-y_1)^{2\beta}} \; .
\end{multline}
The integral over $\nu$ in the right hand side of (\ref{cor12})
for $\beta=1$ and even $D >2$ can be evaluated explicitly
and gives a linear combination of $\zeta$-values with
rational coefficients.
We will publish the explicit formula for (\ref{cor12}) elsewhere.
Here we consider only one special case $\beta=1$ and $D=4$.
%In the case $D=4$ and $\beta=1$
In this case, we have
 $\alpha=\frac{n+1}{2}-i\nu$
 and the master integral~(\ref{cor13})
 is simplified
\begin{align}
 \lb{cor14}
 %\begin{array}{c}
 \int d^4 x_1& d^4 y_2 \, d^4 x\,
\frac{\langle x_1,x_2 |\Psi^{\mu_1\cdots\mu_n}_{\nu,x}\rangle
\langle \Psi^{\mu_1\cdots\mu_n}_{\nu,x} |y_2,y_1 \rangle}{
(x_1-x_2)^{2}(y_1-y_2)^{2}}
 = \nonumber \\ &=(-1)^n \frac{(n+1)}{2^n} \, \tau^3(\nu,n)
\frac{1}{(x_2-y_1)^2} \; ,
 %\end{array}
\end{align}
 where
 $\langle x_1,x_2 |\Psi^{\mu_1\cdots\mu_n}_{\nu,x}\rangle :=
 \Psi^{\mu_1\cdots\mu_n}_{\nu,\beta,x}(x_1,x_2)
 |_{D=4,\beta=1}$ and (see (\ref{cor11}))
 \be
 \lb{cor15}
 \tau(\nu,n) := \left. \tau(\alpha,\beta,n)\right|_{D=4,\beta=1} =
 \frac{(-1)^n (2\pi)^2}{(1+n)^2+4\nu^2} \; .
 \ee
 The coefficient $C_1$ in (\ref{C1}) for $D=4$ %and $\beta=1$
 is reduced to
 \be
 \lb{cor16}
 C_1(n,\nu) =
 \frac{\pi^5}{2^{n+3}(1+n)\, \nu^2}\, \tau(\nu,n)  \; .
 \ee
 Finally, we substitute (\ref{cor14}) -- (\ref{cor16})
 into (\ref{cor12}) and obtain

\begin{align}\label{cor17}
G^{(M)}_2&(x_2,y_1)|_{_{D=4,\beta=1}}  = \nonumber\\ 
 %=\frac{1}{(x_2-y_1)^2} \frac{2^3}{\pi^5}
 %\sum\limits_{n=0}^{\infty}(-1)^n\, (n+1)^2\,
 %\int\limits_{0}^{\infty} d \nu \,\nu^2\, (\tau(\nu,n))^{M+2} = \\
= &\frac{(2\pi)^{2(M+2)}}{(x_2-y_1)^2} \frac{2^3}{\pi^5}
  \sum\limits_{n=0}^{\infty}(-1)^{n(M+1)} (n+1)^2
\times\nonumber\\ \times&\int\limits_{0}^{\infty} d \nu \,
\frac{\nu^2}{\left((1+n)^2+4\nu^2\right)^{M+2}} \nonumber = \\
= &\,\frac{4 \pi^{2M}}{(x_2-y_1)^2} \, C_M  \,
%\frac{4 \pi^{2M}}{(M+1)}\binom{2M}{M}
\sum\limits_{n=0}^{\infty}(-1)^{n(M+1)}
\frac{1}{(n+1)^{2M-1}} \; ,
 %= \\= \frac{4 \pi^{2M}}{(x_2-y_1)^2} C_M
 %\sum\limits_{p=1}^{\infty}(-1)^{(p-1)(M+1)}
 %\frac{1}{p^{2(M+1)-3}} \; ,
\end{align}

where $C_M = \frac{1}{(M+1)} \binom{2M}{M}$ is the Catalan number.
In the last equality in (\ref{cor17}) we used the integral
\begin{align}
\int_{0}^{+\infty} &d\nu\,
\frac{\nu^2}{\left(4\nu^2+\left(1+n\right)^2\right)^{M+2}} = \nonumber\\ =
&\frac{1}{2^5(n+1)^{2M+1}}\,
\frac{\Gamma\left(\frac{1}{2}\right)\Gamma\left(M+\frac{1}{2}\right)}
{\Gamma\left(M+2\right)}
\end{align}
and %take onto account
applied the identity
$\frac{\Gamma(M+\frac{1}{2})\Gamma(\frac{1}{2})}{\Gamma(M+2)}
= \frac{(2M)!\; \pi}{2^{2M} M!(M+1)!}$. Relation
(\ref{cor17}) is equivalent to (\ref{BKr}), (\ref{BK}).
Thus, we have derived the Broadhurst and Kreimer formula
\cite{BrKr}.

Finally, we note
that D.Broadhurst and D.Kreimer fixed
in their paper \cite{BrKr}
the loop measure for each integration over
loop momenta $k$ as $\frac{d^4 k}{\pi^2}$.
Expression (\ref{cor17}) is related
to the $M$ loop zig-zag diagram
(it corresponds to the $(M+1)$ loop contribution to
the $\beta$-function of the $\phi^4_{D=4}$ theory). Therefore,
we have to divide our answer in (\ref{cor17})
by $(\pi^{2})^M$.
In this case, our result (\ref{cor17}) justifies the
normalization factor $(\pi^{2})^M$ in relation (\ref{BKr})
which together with (\ref{BK})
states the Broadhurst and Kreimer conjecture \cite{BrKr}.

\section{Conclusions}
In this letter, we demonstrated how the recent progress in
the investigations
 of the multidimensional conformal field theories
(CFT) can be applied, e.g., in the analytical
evaluations of massless Feynman diagrams. We believe
that the approach described here gives the universal method
of the evaluation of contributions into correlation functions
and critical exponents in various CFT.
We also wonder if it is possible to apply our
$D$-dimensional
generalizations to evaluation similar 4-points 
functions (with fermions)
 that arise \cite{KOP}
 in the generalized ''fishnet'' model, in
 double scaling limit of $\gamma$-deformed $N = 4$ SYM theory.

\section*{Acknowledgements}

   We thank G.Arutyunov, M.Kompaniets and V.Kazakov
 for valuable discussions. This work is supported by the Russian 
 Science Foundation project No 19-11-00131. API is grateful
     to L.Euler International Mathematical Institute in Saint
Petersburg for kind hospitality.

\bibliographystyle{elsarticle-num}
\bibliography{bibl}% Produces the bibliography via BibTeX.

\end{document}